# Structure, energetic and tribological properties and possible applications in NEMS of argon-separated double-layer graphene

*Andrey M. Popov[1,\*], Irina V. Lebedeva[2,\*], Andrey A. Knizhnik[2,3], Yurii E. Lozovik[1,4] and Boris V. Potapkin[2,3]*

[1]Institute of Spectroscopy, Russian Academy of Science, Fizicheskaya Street 5, Troitsk, Moscow, 142190, Russia

[2]Kintech Lab Ltd., Kurchatov Square 1, Moscow, 123182, Russia,

[3]National Research Centre "Kurchatov Institute", Kurchatov Square 1, Moscow, 123182, Russia,

[4]Moscow Institute of Physics and Technology, Institutskii pereulok 9, Dolgoprudny, Moscow Region, 141700, Russia,

*Address correspondence to lebedeva@kintechlab.com, popov-isan@mail.ru

**ABSTRACT** The possibility to control the commensurability and distance between graphene layers separated by a dielectric spacer is considered by the example of a heterostructure consisting of double-layer graphene separated by atomic layers of argon. Van der Waals corrected density functional theory (DFT-D) is applied to study structural, energetic and tribological characteristics of this heterostructure. It is found that in the ground state, monolayer and bilayer argon spacers are incommensurate with the graphene layers, whereas submonolayer argon spacers which are commensurate with the graphene layers can exist only as metastable states. The calculations show that this incommensurability provides negligibly small static friction for relative motion of argon-separated graphene layers and, therefore, such a heterostructure holds great promise for the use in nanoelectromechanical systems. A scheme and operational principles of the nanorelay based on the revealed tribological properties of the heterostructure are proposed.

1. **INTRODUCTION**

Since the discovery of graphene[1] a considerable interest has arisen to various graphene-based nanometer-scale systems. One of the most interesting systems realized recently is double-layer graphene,[2–7] the system consisting of two graphene layers separated by a dielectric spacer. Contrary to bilayer graphene where the layers are placed at the equilibrium interlayer distance of about 3.4 Å, which is close to that in graphite, the distance between the layers in double-layer graphene is determined by the thickness of the dielectric layer separating the graphene layers.

The double-layer graphene has attracted significant theoretical and practical interest. A graphene-based field-effect transistor consisting of two graphene layers with the nanometer-scale distance between the layers was recently implemented.[2,3] Tunable metal-insulator transition was observed in double-layer graphene heterostructures.[4] Electron tunneling between graphene layers separated by an ultrathin boron nitride barrier was investigated.[5] Double-layer graphene heterostructures were also used to determine Fermi energy, Fermi velocity and Landau level broadering.[6] Measurements of Coulomb drag of massless fermions in the double-layer graphene heterostructures were reported[7] and the theory of this phenomenon was considered.[8–14] Theoretical studies of electron-hole pairs condensation in a double-layer graphene were presented.[15–24] Several types of double-layer graphene heterostructures with different dielectric spacers between the layers have been implemented up to now. Namely, graphene layers can be separated by a few-nanometer $Al_2O_3$[6] or $SiO_2$[7] spacer, by one[3,5] or several[3–5] atomic boron nitride layers, and by a layer of adsorbed molecules.[2]

In spite of the essential practical interest, a set of experimental realizations and a considerable body of works devoted to electronic properties of double-layer graphene heterostructures, their structure, energetics and tribological properties are poorly understood. The calculations showed that electronic structure of twisted bilayer graphene changes considerably with changing the twist angle.[25] The tunneling conductance between layers of bilayer graphene changes by a few times upon relative displacement of the layers[26,27] and by an order of magnitude upon relative rotation of layers.[26] The electronic properties of double-layer graphene with a thin dielectric spacer can also depend significantly on the relative position and orientation of the layers as well as on the interlayer distance. Thus, production of double-layer graphene with the controllable commensurability of the graphene layers and distance between the layers is both of fundamental and practical interest. Since $Al_2O_3$[6] or $SiO_2$[7] spacers are not layered materials, they do not allow implementation of double-layer graphene with a predetermined interlayer distance. As for boron nitride, the lattice constant of this material is only 2% greater than the lattice constant of graphene.[28] Thus, the commensurate-incommensurate phase transition should be expected for such a nearly commensurate system and should lead to a pattern of alternate commensurate and incommensurate regions so that the same relative position of layers in the whole system is not possible (see Ref. 29 for bilayer graphene with one stretched layer). We propose here that the layers of adsorbed molecules on graphene surface can be used to obtain double-layer graphene both with the controllable commensurability and distance between the layers. This possibility is studied by the example of argon-separated double-layer graphene.

In the last decade, a wide set of nanoelectromechanical systems (NEMS) based on the free relative motion of nanotube walls has been proposed and implemented (see Ref. 30 for a review). Based on the recent experimental studies of relative motion of atomically smooth graphene layers, it has been suggested that analogous NEMS based on this motion can be also elaborated.[31] However, it has turned out that due to commensurability of graphene layers in the usual graphene bilayer, the barriers to their relative motion are not sufficiently low[32–36] and strongly influence dynamic tribological characteristics of the system. As a result, many of the proposed NEMS based on relative in-plane motion of graphene layers cannot be realized.[37] Thus, possible applications of interaction between graphene layers in NEMS proposed to date are restricted to a nanoresonator based on sub-Angstrom relative in-plane vibrations of graphene layers,[38] a force sensor based on sub-Angstrom relative in-plane displacement of graphene layers,[27] a force sensor and memory cell based on changes of the distance between the layers.[39] Here we show that the presence of adsorbed molecules between graphene layers can lead to disappearance of the barriers to relative motion of the layers and makes possible development of NEMS based on such a motion.

We apply the van der Waals corrected density functional theory (DFT-D) to calculate the structural, energetic and tribological characteristics of double-layer graphene separated by commensurate and incommensurate atomic layers of argon. Advantages of tribological characteristics of the considered heterostructure for elaboration of NEMS are discussed.

## 2. METHODS

Zero-temperature calculations of potential energy for graphene layers separated by argon atoms are performed in the framework of dispersion-corrected density functional theory (DFT-D) using VASP code[40] with the generalized gradient approximation (GGA) density functional of Perdew, Burke, and Ernzerhof[41] corrected with the dispersion term (PBE-D).[42] The basis set consists of plane waves with the maximum kinetic energy of 500 eV. The interaction of valence electrons with atomic cores is described using the projector augmented-wave method (PAW).[43] A second-order Methfessel–Paxton smearing[44] with a width of 0.1 eV is applied.

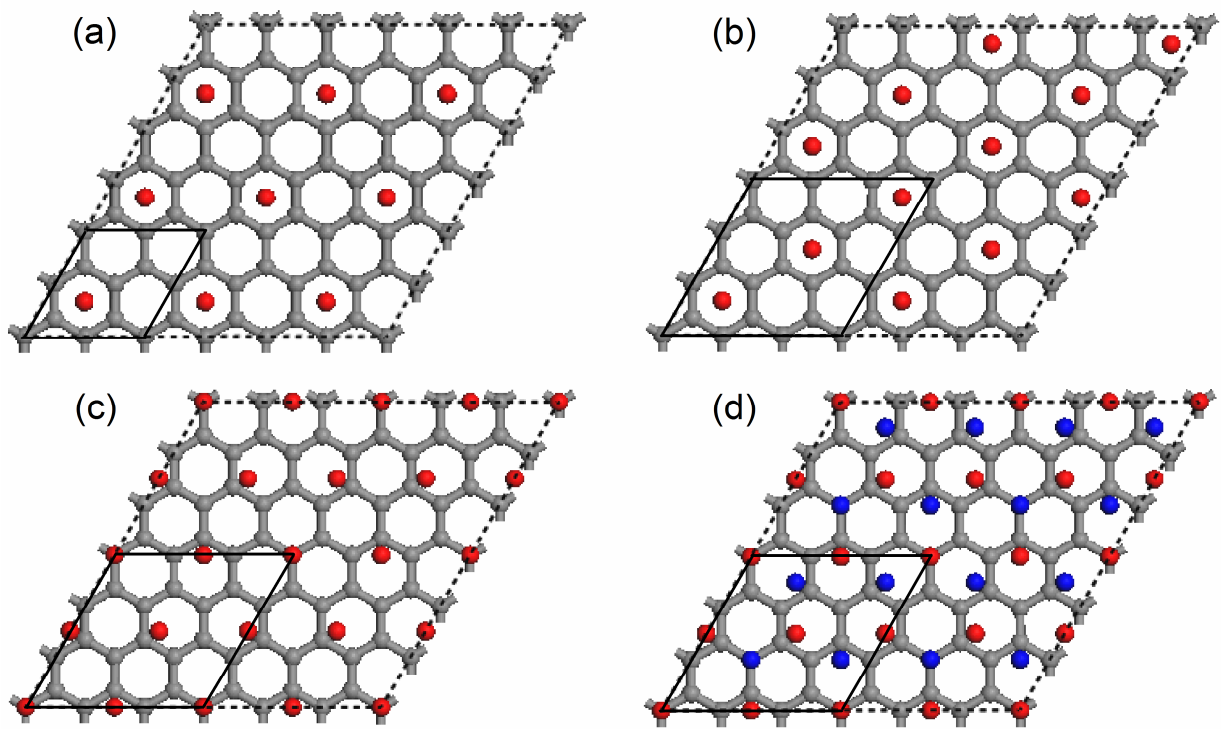

**Figure 1.** Structures of the argon-separated double-layer graphene with different argon spacers: (a) spacer A, (b) spacer B, (c) spacer C and (d) spacer D. Only one graphene layer is shown for clarity. Carbon atoms are coloured in gray. Argon atoms of the first and second atomic layers are coloured in red and blue, respectively. Model cells are indicated by solid lines.

Four argon spacers of different structure are considered. Spacers A and B are submonolayer argon layers commensurate with the graphene layers in which the periods of argon lattice are multiples of the lattice constant of graphene (Figure 1a and b). Argon to carbon ratios in the double-layer graphene with the spacers A and B are Ar:C = 1:16 and 1:12, respectively. In addition, since we are not able to simulate incommensurate argon-graphene heterostructures under the periodic boundary conditions (PBCs) directly, we consider spacers C and D with inequivalent positions of argon atoms on the graphene lattice within a model cell (Figure 1c and d). The results below show that only 4 argon atoms with nonequivalent positions for spacer C are sufficient for the dramatic change of tribological characteristics of the system. Thus, spacers C and D can be considered as prototypes of the spacers which are incommensurate with the graphene layers. Argon to carbon ratios in double-layer graphene with the spacers C and D are Ar:C = 1:9 and 2:9, respectively. In these spacers, the period of argon lattice equals one and a half lattice constants of graphene. This is only 2% smaller than the equilibrium distance between atoms in the isolated argon layer of 3.78 Å obtained in our calculations, which is close to the results of experimental observations.[45,46] Therefore, argon coverages for the spacers C and D are very close to the monolayer and two atomic layers of argon, respectively.

We do not consider here the possibility of symmetry distortions of the argon spacers A and B (such a consideration would need the choice of PBCs compatible with each certain symmetry distortion) and treat them as rigid. The negligibly small barriers are found below for relative motion of graphene layers separated by the spacers C and D with the rigid structure. The distortions of the spacer structure can only decrease these barriers and thus do not affect this qualitative result. It should also be noted that though commensurate krypton layers on graphite[45–48] and lateral surfaces of carbon nanotubes[49] are well studied, there is no evidence for symmetry distortions.

An oblique-angled simulation cell is considered. The model cell has two equal sides at an angle of 60° and a perpendicular side of 20 Å length. The length of two equal sides is 4.932 Å for the spacer A and 7.398 Å for the spacers B, C and D. Integration over the Brillouin zone is performed using the Monkhorst-Pack method[50] with 18 x 18 x 1 and 12 x 12 x 1 k-point sampling, respectively. The structures of the graphene layers are separately geometrically optimized using the conjugated gradient method until the residual force acting on each atom becomes less than 0.01 eV/Å. After that the structures of the graphene layers are considered as rigid. Account of structure deformation induced by the interlayer interaction was shown to be inessential for the shape of the potential relief of interaction energy between graphene-like layers, such as the interwall interaction of carbon nanotubes[51] and the intershell interaction of carbon nanoparticles.[52, 53]

## 3. RESULTS AND DISCUSSION

We have started our consideration of properties of the argon-separated double-layer graphene from calculation of relative positions of the graphene layers and the argon spacer with the rigid structure corresponding to the minimum of the potential energy of the system. For this purpose the dependences of the potential energy on the in-plane displacements of each of the graphene layers relative to the argon spacer and the distances between the graphene layers and the argon spacer have been calculated. The contributions of interaction between the graphene layers, between each of the graphene layers and the argon spacer and inside the argon spacer to the total interaction energy of argon-separated double-layer graphene (the interactions inside the graphene layers are excluded from this quantity) for the found minimum energy positions have been also obtained. To evaluate these contributions the energies of the systems consisting of two graphene layers at the same interlayer distance and in-plane relative position as in the argon-separated double-layer graphene but without the argon spacer, the isolated graphene layer and the isolated argon spacer have been calculated. The results of these calculations for all the considered structures of the argon spacer are listed in Table 1.

The potential reliefs, i.e. the dependences of interaction energy between the graphene layer and argon spacer on coordinates describing their relative in-plane displacements, are also calculated for the argon-graphene distance that is equilibrium for the double-layer graphene. The magnitudes of corrugation of the potential energy relief and barriers to translational motion of each of the graphene layers relative to the argon layer are presented in Table 1. For the spacers A and B, the minima of these potential reliefs correspond to the relative positions of the spacer and graphene layers in which argon atoms are placed in the centers of hexagons of the graphene lattice. For the spacers C and D, the corrugations of these potential reliefs are too small to determine the relative position of the spacer and graphene layers corresponding to the energy minimum.

The calculated equilibrium distance for graphene layers separated by the spacers A, B, and C with the monolayer or less argon coverage lies in the range from 6.6 to 6.8 Å (Table 1), in qualitative agreement with the experimental value of ~6 Å for undetermined adsorbed molecules.[2] In the case of the spacer D with two atomic argon layers, the distance between the graphene layers increases to 10 Å. The values of interaction energies between the graphene layers and between the graphene layers and the argon spacer (Table 1) show that at these equilibrium distances, the coupling of the graphene layers is provided mainly via graphene-argon interactions, whereas for the spacer D, the contribution of the graphene-graphene interaction into this coupling is completely negligible.

**Table 1**. Calculated equilibrium distances $d_{gr-gr}$ and $d_{gr-Ar}$ between the graphene layers and between each graphene layer and the argon layer, total interaction energy $E_b$, interaction energy $E_{Ar-gr}$ between the argon and graphene layers, interaction energy $E_{gr-gr}$ between the graphene layers, interaction energy $E_{Ar-Ar}$ between argon atoms and barrier $\Delta E_{gr}$ to relative motion of the graphene layers per carbon atom of one of the graphene layers for the double-layer graphene with different structure of the argon spacer. Calculated magnitude $\Delta E_{max}$ of corrugation of the potential energy relief and barrier $\Delta E_{tr}$ to relative motion of a single graphene layer relative to the argon spacer per carbon atom of the graphene layer for the indicated distance $d_{gr-Ar}$ are also given. The structures of argon-separated double-layer graphene are shown in Figure 1. The data for argon-free bilayer graphene are taken from paper[34] for reference.

| Ar:C | $d_{gr-gr}$ (Å) | $d_{gr-Ar}$ (Å) | $E_b$ (meV) | $E_{gr-Ar}$ (meV) | $E_{gr-gr}$ (meV) | $E_{Ar-Ar}$ (meV) | $\Delta E_{max}$ (meV) | $\Delta E_{tr} = \Delta E_{gr}$ (meV) |
|---|---|---|---|---|---|---|---|---|
| 1:16 | 6.59 | 3.30 | -30.2 | -23.4 | -4.4 | -2.41 | 1.23 | 1.09 |
| 1:12 | 6.64 | 3.32 | -40.6 | -28.5 | -4.3 | -7.84 | 1.47 | 1.29 |
| 1:9 | 6.82 | 3.41 | -51.2 | -34.0 | -3.9 | -13.3 | <6·10⁻³ | <6·10⁻³ |
| 2:9 | 9.97 | 3.42 | -74.2 | -37.3 | -1.4 | -35.4 | <6·10⁻³ | <6·10⁻³ |
| Ar-free bilayer[34] | 3.25 | 3.25 | -50.6 | | -50.6 | | 19.5 | 2.1 |

To estimate the contribution of graphene-graphene interaction to variations of the potential energy of the system upon relative in-plane displacements of the graphene layers we have performed calculations of the potential energy relief for two graphene layers at the same distance as in the argon-separated double-layer graphene but without the argon spacer. The magnitudes of corrugation of these potential reliefs are found to be below 0.003 meV per carbon atom of one of the layers, which is more than two orders of magnitude smaller than the magnitudes of corrugation calculated for the graphene-argon interaction for the spacers A and B (Table 1). Therefore, the graphene-graphene interaction provides a negligibly small contribution to the variations of the potential energy upon relative motion of graphene layers. As a result, in commensurate systems, the AA-stacking of graphene layers with the equivalent positions of argon atoms in the centers of hexagons of the both graphene lattices is observed to be the most energetically favorable.

The total interaction energy of the argon-separated double-layer graphene appears to increase in magnitude with increasing the argon coverage, i.e. is higher in magnitude for the C and D spacers serving as prototypes of incommensurate states of the argon-separated double-layer graphene in our calculations compared to the commensurate A and B spacers. No commensurate phase of argon on graphite[45–48] or lateral surfaces of carbon nanotubes[49] has been observed experimentally. However, for argon on graphite surface, the argon-graphene interaction energy is twice smaller in magnitude compared to the double-layer graphene and thus commensurate argon layers between graphene layers are more stable than on graphite and cannot be excluded as metastable states. Nevertheless, for the commensurate A and B spacers, the total interaction energy of the double-layer graphene is found to be smaller in magnitude than the interaction energy between graphene layers in the argon-free bilayer, while for the systems with the spacers C and D, the opposite relation takes place (Table 1). Therefore, let us consider stability of the double-layer graphene with the commensurate spacers A and B with respect to decay into regions with the incommensurate argon spacer (with argon coverage close to that of the spacer C) and argon-free regions. Upon the decay of the commensurate spacers A and B, these argon-free regions should occupy up to approximately 7/16 and 1/4 of the graphene area, respectively. The analysis of calculated total interaction energies for the argon-separated double-layer graphene, argon-free double-layer graphene with the same interlayer distance and bilayer graphene (which can form in result of sticking of the graphene layers) shows that this decay should occur only in the case where graphene layers stick together in the argon-free regions. Due to the high flexibility of graphene layers, this decay can easily take place in heterogeneous systems where the argon-free regions with stuck graphene layers are already present. However, in the case of homogeneous commensurate argon spacer, the critical nucleus size of the argon-free regions with stuck graphene layers can be too large for thermally activated decay, providing the possibility to observe the double-layer graphene with the commensurate argon spacer as a metastable state. As opposed to argon, krypton on graphite[54–56] and lateral surfaces of carbon nanotubes[49] is known to form a commensurate phase at submonolayer coverages and low temperatures. Therefore, we predict that the krypton-separated double-layer graphene can be observed as a commensurate system.

The barriers to relative motion of the graphene layers in the argon-separated double-layer graphene can be extracted from the analysis of the potential relief of interaction energy of the argon spacer with each of the graphene layers at the argon-graphene distance which is equilibrium for the double-layer graphene. The calculations within the DFT-D approach show that the shape of the potential relief of interaction energy between the graphene layer and the commensurate argon layer can be approximately described by the expression containing only the first Fourier components (Figure 2)

$$U_{\text{gr-Ar}}(x,y,z) = U_1(z)\left(3 - 2\cos(k_1 x)\cos(k_2 y) - \cos(2k_2 y)\right) + U_0(z), \qquad (1)$$

where $k_1 = 2\pi / a_0$, $k_2 = 2\pi / (\sqrt{3} a_0)$, $a_0 = 2.46$ Å is the lattice constant of graphene, $x$ and $y$ are coordinates corresponding to the in-plane relative displacement of argon spacer and graphene layer ($x$ and $y$ axes are chosen along the armchair and zigzag directions, respectively) and $z$ is the distance between the argon spacer and graphene layer. The values of energy parameters are found to be $U_0 = -23.4$ meV and $U_1 = 0.27$ meV per carbon atom for the spacer A and $U_0 = -28.5$ meV and $U_1 = 0.33$ meV per carbon atom for the spacer B. The relative root-mean square deviations $\delta U / U_1$ of the potential energy reliefs obtained using Equation 1 from the potential energy reliefs obtained by the DFT-D calculations are within 2% for both of the spacers A and B. The magnitudes of corrugation of the potential energy reliefs are found to be $\Delta E_{max} \approx 4.5 U_1 = 1.2$ and 1.5 meV per carbon atom for the spacers A and B, respectively. The barriers to relative motion of the graphene layer and the argon layer are calculated to be $\Delta E_{tr} \approx 4 U_1 = 1.1$ and 1.3 meV per carbon atom, respectively. Note also that the first Fourier components were previously shown to be sufficient for the description of the potential reliefs of interlayer interaction energy in bilayer graphene[34–36,38,57] and interwall interaction energy in carbon nanotubes.[51,58,59]

As shown above, the contribution of graphene-graphene interaction into the variation of the total interaction energy upon relative motion of the graphene layers in the double-layer graphene is negligibly small. Therefore, this variation is given by the sum of variations in the interaction energies between the argon layer and each of the graphene layers, both of which can be approximated by Equation 1

$$\delta U_{\text{gr-Ar-gr}}(x_0, y_0, z_0; x, y) = U_1(z_0/2)\left(3 - 2\cos(k_1 x)\cos(k_2 y) - \cos(2 k_2 y)\right) \\ + U_1(z_0/2)\left(3 - 2\cos(k_1(x - x_0))\cos(k_2(y - y_0)) - \cos(2 k_2(y - y_0))\right) \quad (2)$$

The total energy of the system with the relative position $x_0, y_0, z_0$ of graphene layers can be obtained by minimizing this sum with respect to the position of the commensurate argon layer

$$\delta U_{\text{gr-gr}}(x_0, y_0, z_0) = \min_{x,y}\left[\delta U_{\text{gr-Ar-gr}}(x_0, y_0, z_0; x, y)\right]. \quad (3)$$

The potential relief calculated using Equation 3 is given in Figure 3.

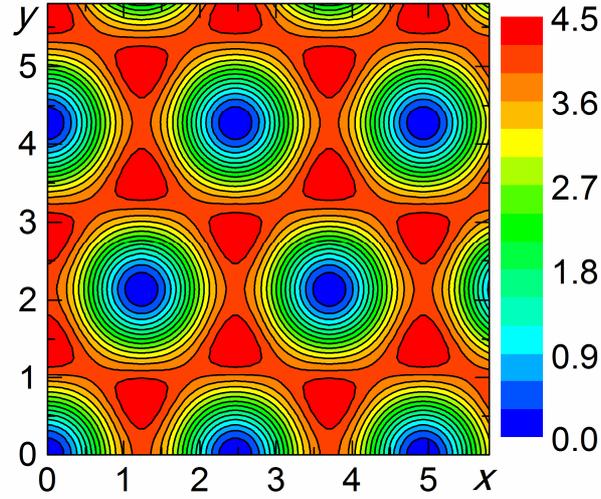

**Figure 2.** Interaction energy (in $U_1$ units; see Equation 1) between the graphene layer and the commensurate argon layer as a function of the relative position of the argon layer $x$ and $y$ (in Å; $x$ and $y$ axes are chosen along the zigzag and armchair directions, respectively) approximated using Equation 1. The position $x = 0$ and $y = 0$ corresponds to the ground state of the commensurate argon layer on the graphene layer.

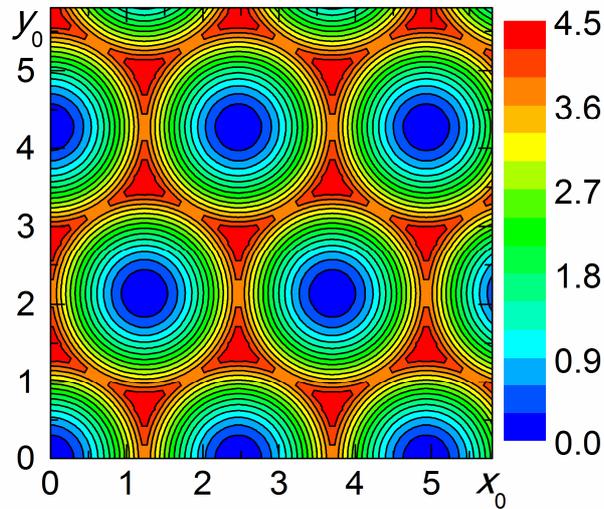

**Figure 3.** Interaction energy (in $U_1$ units; see Equation 1) of graphene layers in the double-layer graphene with the commensurate argon layer as a function of the relative position of the graphene layers $x_0$ and $y_0$ (in Å; $x$ and $y$ axes are chosen along the zigzag and armchair directions, respectively) obtained using Equation 3. The position $x_0 = 0$ and $y_0 = 0$ corresponds to the AA stacking of graphene layers.

Figure 2 shows that the minimum energy path for motion of the commensurate argon layer relative to each of the graphene layers lies along the zigzag direction. So the energy favorable path for relative motion of graphene layers in the argon-separated double-layer graphene corresponds to the case when the graphene layers are displaced along the zigzag direction (see also Figure 3). In this case, the position $x$ of the argon spacer at which the energy of the system is minimized for a given relative position $x_0$ of the graphene layers is determined by the conditions

$$\partial U_{\text{gr-Ar-gr}} / \partial x = 4k_1 U_1 \sin(k_1(x - x_0/2)) \cos(k_1 x_0/2) = 0 \qquad (4)$$

$$\partial^2 U_{\text{gr-Ar-gr}} / \partial^2 x = 4k_1^2 U_1 \cos(k_1(x - x_0/2)) \cos(k_1 x_0/2) \geq 0$$

From these equations, we find that the barrier to relative motion of the graphene layers in the double-layer graphene with the commensurate argon layer is exactly equal to the barrier to relative motion of the argon layer on one of the graphene layers $\Delta E_{\text{gr}} = 4U_1 \approx 1.1$ and 1.3 meV per carbon atom of one of the graphene layers for the spacers A and B, respectively. It is seen that this barrier is twice smaller than that for the relative motion of the graphene layers in the argon-free bilayer (Table 1).

For the argon spacers C and D with inequivant positions of argon atoms within the model cells, the magnitudes of corrugation of the potential energy reliefs are found to be below $6 \cdot 10^{-3}$ meV per carbon atom. This means that for incommensurate states of the argon-separated double-layer graphene, the potential energy relief is extremely smooth and, therefore, the static friction force for relative motion of graphene layers is negligibly small. Thus, tribological properties of the argon-separated double layer graphene are completely different from those of bilayer graphene, where the significant barrier for relative motion of the layers exists for commensurate relative orientations of the layers.[33–36,57] Due to this difference in the tribological characteristics, different dynamic behavior associated with the interaction and relative motion of the layers should be expected for these heterostructures. In particular, the self-retraction of telescopically extended graphene layers separated by the monolayer or bilayer argon spacer can take place for any relative orientations of the layers, whereas for bilayer graphene, such a self-retraction requires the relative rotation of the layers to incommensurate orientations, which has a considerable energy cost.[35,36] In the following section, we consider how these different tribological properties of the argon-separated double-layer graphene can be advantageous for its application in NEMS.

## 4. APPLICATION IN NANOELECTROMECHANICAL SYSTEMS

The negligibly small static friction for relative motion of argon-separated graphene layers shows promise for application of this heterostructure in NEMS based on the interaction and relative motion of graphene layers. This property of the argon-separated double-layer graphene is similar to that of multi-walled carbon nanotubes, for which the relative motion of adjacent walls occurs with the negligibly small static friction force in majority of cases (see Ref. 30 and references therein). Thus, operational principles of NEMS based on the Ar-separated double-layer graphene can be basically the same as operational principles of NEMS based on the interaction and relative motion of nanotubes walls. Examples of such NEMS include variable nanoresistors[60–62] and nanocapacitors[62] based on dependences of conductance and capacity of the system, respectively, on the overlap length of nanotube walls. We propose that analogous NEMS can be based on the dependence of conductance or capacity of the system on the overlap area of graphene layers in the argon-separated double-layer graphene. Such a system can be considered as a nanoresistor or nanocapacitor depending on the tunneling conductivity through the argon spacer. Since the static friction force for relative motion of graphene layers in this system is negligibly small, the overlap area of the layers can be tuned precisely by a piezo-actuator attached to one of the layers.

The negligibly small static friction for relative motion of nanotube walls has been used to implement nanorelays based on the telescopic motion of nanotubes walls.[63,64] Switching on of these nanorelays occurs under the action of the electrostatic force between movable telescopic walls of two coaxial multi-walled nanotubes, whereas switching off takes place in result of self-retraction of these walls back into the nanotubes due to the van der Walls force. By analogy, here we propose a scheme and operational principles of the nanorelay based on the relative rotation of the graphene layers 4 and 5 separated by the argon spacer 6 (Figure 4). The operation of this nanorelay is determined by the balance of torques of the forces that act on the movable graphene layer 5. The force $F_v$ of the van der Waals interaction between the movable graphene layer 5 and the argon spacer 6 leads to retraction of the layer 5 back upon its extension beyond the fixed layer 4. The torque of this force induces rotation of the layer 5 so that the angle α increases (the angle α is shown in Figure 4). The torques of van der Waals forces $F_{a2}$ and $F_{a3}$ and electrostatic forces $F_{e2}$ and $F_{e3}$ of interaction between the movable graphene layer 5 and the electrodes 2 and 3, respectively, provide rotation of the layer 5 so that the angle α decreases. When a voltage is applied between the electrodes 2 and 3, an electric dipole moment is induced in the movable graphene layer 5 and electrostatic forces $F_{e2}$ and $F_{e3}$ between this layer and electrodes 2 and 3 arise. If the total torque of the forces $F_{a2}$, $F_{a3}$, $F_{e2}$ and $F_{e3}$ that attract the movable layer to the electrodes is greater than the torque of the retraction force $F_v$, the nanorelay switches from the non-conducting state OFF to the conducting state ON.

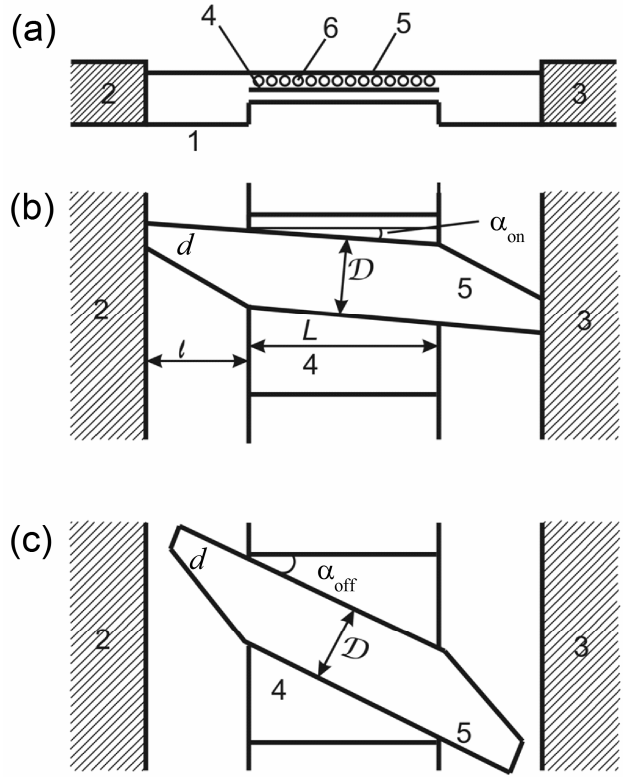

**Figure 4.** The scheme of the nanorelay based on the double-layer graphene with the argon spacer: a) conducting state ON, side view; b) conducting state ON, top view; c) non-conducting state OFF, top view. The dielectric substrate (1), electrodes (2) and (3), fixed and movable graphene layers (4) and (5), respectively, separated by the argon spacer (6) are indicated.

In the absence of applied voltage, the state of the nanorelay is determined by the balance of torques of only the van der Waals forces $F_{a2}$, $F_{a3}$, and $F_v$. In the case where the torque of retraction force $F_v$ is less than the total torque of forces $F_{a2}$ and $F_{a3}$, the nanorelay can remain in state ON in the absence of applied voltage. In this case, the nanorelay can be used as a nonvolatile memory cell for long-term data storage. Otherwise, the nanorelay can be used as a volatile memory cell for primary data storage. It should be noted that the width of the fixed layer 4 should be less than the length of the movable layer 5 to prevent the total retraction of the layer 5.

Let us discuss advantages of the proposed nanorelay in comparison with the nanorelays based on the relative motion of telescopic nanotubes walls. For the nanorelays based on the relative motion of telescopic nanotubes walls, both the retraction force and the force of interaction between movable telescopic walls are determined by a single geometric parameter, the diameter of the movable wall. Thus, designing such a nanorelay with a pre-assigned characteristic to be nonvolatile or volatile memory cell presents a real challenge. For the proposed memory cell based on the relative motion of argon-separated graphene layers, the forces determining this property depend on different geometric parameters of the movable layer. Namely, the retraction force $F_v$ is determined by the width $D$ of the central part of movable layer, whereas the van der Waals forces $F_{a2}$ and $F_{a3}$ of interaction between the movable graphene layer 5 and the electrodes 2 and 3 are determined by the width $d$ of outermost part of the layer 5 (Figure 4). Moreover, the arms of these forces are determined also by two parameters. Namely, the arm of the force $F_v$ is equal to $L/2$, where $L$ is the length of the fixed layer 4, whereas the arm of the forces $F_{a2}$ and $F_{a3}$ is equal to $L/2+l$ where $l$ is the distance between the fixed layer 4 and the electrodes 2 and 3. Thus, the operational characteristics of the proposed nanorelay are determined by four different geometric parameters. The appropriate choice of these parameters makes possible to design nanorelays which can operate as nonvolatile or volatile memory cells.

Carbon nanotube-based systems typically have rather high values of Q-factors, which were measured to be about 500 for bending oscillations of nanotubes[65] and were predicted to lie in the range of 100–1000 for telescopic oscillations of nanotubes walls (see Ref. 66, 67 and references therein). Mechanical oscillations (followed by the oscillation of the tunneling current) arising after switching on were shown to be the main problem which restricts the response speed of memory cells based on bending of nanotubes[68] and relative telescopic motion of nanotubes walls.[69] Molecular dynamics simulations show that, contrary to nanotubes, relative telescopic oscillations of graphene layers are suppressed because of high dynamic friction related to the excitation of flexural vibrations of the flake.[37] Such vibrations should be excited also during the relative telescopic motion of the Ar-separated graphene-layers. Thus, we assume that proposed nanorelay can be used as a memory cell with a high operational frequency.

## 5. CONCLUSION

We consider here the possibility to use layers of adsorbed molecules on graphene surface for production of double-layer graphene with the controllable commensurability of the graphene layers and distance between the layers. These heterostructures show considerable promise both for studies of fundamental phenomena and for the use in nanoelectronics. We have applied the van der Waals corrected density functional theory for calculations of the structural, energetic and tribological characteristics of the heterostructure consisting of two graphene layers separated by the submonolayer, monolayer and bilayer argon spacers. The contributions of interaction between the argon atoms, between the argon spacer and graphene layers, and between the graphene layers into the total interaction energy of the system have been found for different argon coverages. The analysis of calculated energies has shown that the heterostructures consisting of two graphene layers separated by the commensurate submonolayer argon spacer can decay into the regions with monolayer argon coverage where the argon layer is incommensurate with the graphene layers and the regions free of argon where the graphene layers are stuck. Therefore, the submonolayer argon spacers commensurate with the graphene layers can exist only as metastable states. The double-layer graphene separated by the incommensurate monolayer or bilayer argon spacer is the stable system with the distance between the graphene layers which has been calculated to be 6.82 Å and 9.97 Å for the monolayer and bilayer argon spacer, respectively. We believe that the double-layer graphene in which the commensurate graphene layers are separated by the commensurate krypton spacer should be a stable system.

The potential relief of interaction energy between the argon spacer and a graphene layer at the constant argon-graphene distance has been calculated for different argon coverages. Based on these calculations, it has been shown that the barrier (and, therefore, the static friction force) for relative motion of graphene layers separated by the incommensurate monolayer or bilayer argon spacer is negligibly small for any relative orientations of the layers. This result means that the tribological characteristics of the considered heterostructure differ from those of the bilayer graphene,[35–37] where the significant barrier for relative motion of the layers exists for commensurate relative orientations of the layers. Due to this difference in the tribological characteristics, the difference in the behavior associated with the interaction and relative motion of the layers should be expected. In particular, contrary to bilayer graphene, the self-retraction of telescopically extended graphene layers separated by the monolayer or bilayer argon spacer can easily take place for any relative orientations of the layers. For the highest coverage possible for the commensurate argon spacer, the barrier to relative motion of graphene layers is found to be 1.3 meV per carbon atom of one of the graphene layers. This is less than the value of this barrier for the bilayer graphene about 2 meV[34] obtained also in the framework of DFT-D using VASP code.

The revealed tribological characteristics of double-layer graphene separated by the incommensurate argon spacer allow us to assume that such heterostructures are promising for applications in nanoelectromechanical systems. These applications can include variable nanocapacitors or nanoresistors depending on the conductivity between argon-separated layers and nanorelays based on relative motion or rotation of such layers (such nanorelays can be used also as memory cells). Here we propose the scheme and operational principles of the nanorelay based on relative rotation of argon-separated graphene layers. The operational switching frequency of nanotube-based nanorelays is restricted by mechanical oscillations of the movable part of the nanorelay (and related oscillations of the current) after switching on.[68,69] The Q-factor for relative in-plane oscillations of graphene layers[37] ( $Q \sim 1$ ) was found to be several orders of magnitude less than the Q-factors of nanotube-based systems. Thus, a high value of operational switching frequency can be expected for the proposed graphene-based nanorelay.

Let us also discuss the current state of art in production of graphene-based systems. In the pioneering work of Novoselov *et al.*, graphene flakes were placed on an insulating substrate and brought into contact with electrodes[1] (to create a field-effect transistor). A further considerable progress has been achieved in manipulation of individual graphene layers. The possibilities to cut graphene nanoribbons with desirable geometrical parameters[70] and remove individual graphene layers in a controllable way for device patterning[71] were demonstrated. All these give us a cause for optimism that the proposed heterostructures and nanoelectromechanical systems based on the argon-separated double-layer graphene will be implemented in the near future.

**ACKNOWLEDGEMENTS**

This work has been supported by the RFBR grants 11-02-00604 and 12-02-900241-Bel_a and Samsung Global Research Outreach Program. The atomistic calculations are performed on the Supercomputing Center of Lomonosov Moscow State University[72] and on the Multipurpose Computing Complex NRC "Kurchatov Institute".[73]

(73) http://computing.kiae.ru/